# Contributions of Co and Fe orbitals to Perpendicular Magnetic Anisotropy of MgO/CoFeB Bilayers with Spin-Orbit-Torque-Related (Ta, W, IrMn, and Ti) Underlayers

Sanghoon Kim[1,†], Seung-heon Chris Baek[2,3], Mio Ishibashi[1], Kihiro Yamada[1], Takuya Taniguchi[1], Takaya Okuno[1], Yoshinori Kotani[4], Tetsuya Nakamura[4], Kab-Jin Kim[1,5], Takahiro Moriyama[1], Byong-Guk Park[2], and Teruo Ono[1,†]

[1]Institute for Chemical Research, Kyoto University, Uji, Kyoto 611-0011, Japan

[2]Department of Materials Science and Engineering, KAIST, Daejeon 34141, Republic of Korea

[3]School of Electrical Engineering, KAIST, Daejeon 34141, Republic of Korea

[4]Japan Synchrotron Radiation Research Institute (JASRI), Sayo, Hyogo 679-5198, Japan

[5]Department of Physics, Korea Advanced Institute of Science and Technology, Daejeon 34141, Korea

† Correspondence to: makuny80@gmail.com, ono@scl.kyoto-u.ac.jp

Abstract

We study the perpendicular magnetic anisotropy (PMA) of the CoFeB/MgO bilayers in contact with W, Ta, IrMn and Ti which has been suggested as the spin-orbit-torque-related underlayers. The saturation magnetization of the CoFeB depends on the underlayer materials due to formation of a dead-layer, affecting PMA strength of each film. The x-ray magnetic circular dichroism measurement reveals that the interfacial intermixing suppresses only the perpendicular orbital moment of Fe, while the intermixing simultaneously suppresses both the perpendicular and in-plane orbital moments of Co.

As there are immense demands for increase in storage capacity of the magnetoelectric devices such as magnetic tunnel junctions (MTJ), hard disk drive and spin-logic devices, thermal stability is one of the most considerable parameters [1-7]. Therefore, it is required to use ferromagnetic (FM) thin-films having strong perpendicular magnetic anisotropy (PMA) to overcome the thermal instability arising from size reduction of such devices. Multilayer systems based on a CoFeB/MgO junction have been intensively studied because both strong PMA and high spin polarized tunneling current can be achievable. The spin polarization of the tunneling current can be higher than 70% with a [001]-oriented MgO tunnel barrier, which can drastically increase tunneling magnetoresistance up to several hundred percent [8-10]. Note that the highly [001]-textured MgO tunnel barrier can be formed with an amorphous CoFeB by using the magnetron sputtering method [7, 11]. It has been also discovered that recrystallization of CoFeB during annealing process gives rise to sizable PMA of the MgO-based MTJs in order of $10^5$ J/m$^3$ [5,12]. Thus, the CoFeB/MgO junction has been an important ingredient for revealing high performance spin devices with strong thermal stability.

In recent years, spin-orbit torque (SOT)-induced magnetization switching has been proposed as an efficient way for operation of the spintronic devices [13,14]. The use of a non-magnetic metal (NM) electrode which has the large spin Hall angle is crucial for operational efficiency of the SOT-based device [15-21]. In these respects, it is inevitable to combine SOT-related NM electrodes with the CoFeB/MgO junction. Here, we examine PMA of CoFeB/MgO bilayers with W, Ta, Ti, and IrMn underlayers which has been recently suggested as the electrode of the SOT-based devices [13-21]. Our x-ray absorption spectroscopy (XAS) and x-ray magnetic dichroism (XMCD) study manifest the role of the Co and Fe orbitals for the PMA with those underlayers.

Four different $Co_{32}Fe_{48}B_{20}$ (termed CoFeB)-based films, listed in Table I, were prepared using d.c. and a.c. sputtering methods on a thermally oxidized Si substrate. The base pressure was under $4.0 \times 10^{-6}$ Pa. A working pressures for a.c. and d.c. sputtering were 1.33 Pa and 0.40 Pa, respectively. As-deposited films were annealed at 150°C for 30 minutes to form PMA. The remanence magnetization curves were measured by a superconducting quantum interference device (SQUID). The soft XAS at both the Co and Fe $L$ edges were measured using the total electron yield method with 96% circularly polarized incident x-rays under an applied magnetic field of 1.9 T. Measurements were performed at the BL25SU beam-line in SPring8. To evaluate the anisotropy of orbital magnetic moments ($m_o$), both 0° and 70° incident angles of x-rays with respect to the film normal were used.

Figure 1 presents the remanence magnetization curves, which show that all of the films exhibit PMA. It has been reported that the $M_s$ of CoFeB with PMA is 1.00–1.25 MA/m after the annealing process [8,22,23]. However, the films with Ti, Ta, and IrMn underlayers exhibit lower $M_s$ values than the reported value. In particular, the $M_s$ values of films with Ti and Ta were 0.61 and 0.47 MA/m, respectively. These results show that the CoFeB layers have a magnetically dead layer, which is especially severe with Ta and Ti [24,25]. Intermixing between CoFeB and seed layers during thermal annealing or oxidization at the interface with MgO are possible causes of the dead layer. The B diffusion effect during annealing on the dead layer is negligible because it does not affect the $M_s$ of CoFeB [23,26]. In this study, the dead layers in our films are mainly formed from intermixing between CoFeB and the underlayers because there was no significant oxidization of Co or Fe in our films according to the XAS study. This will be discussed later.

Considering that PMA arises mainly from the CoFeB/MgO interface ($K_i$), $K_i$ in each film is estimated using the relation: $B_s = B_a - \mu_0 M_s$, where $B_s$ is a saturation field with hard axis (indicated with red arrows in Fig. 1 (a)-(d)), $B_a$ is an anisotropy field ($B_a = 2K_i/t_{CoFeB} \cdot M_s$). Measured $M_s$ and $B_s$ values of the films are listed in Table I. The estimated $K_i$ values show clear underlayer-dependence in a range between ~0.2-0.9 mJ/m$^2$ as plotted in Fig. 1 (e). The trend of $K_i$ is consistent with that of $M_s$ value, while $B_s$ does not follow the trend.

The XAS measurement was performed to observe how the chemical state at the CoFeB/MgO interface is influenced by the underlayer. Figure 2 shows XAS spectra at the Co and Fe $L$ edges. All XAS spectra of Co and Fe exhibit typical metallic peaks at both $L_3$ and $L_2$ edges, which means that there are no clear oxide phases. The first-derivative XAS can be used to clarify the chemical states clearly [27]. As shown in Figs. 2(c) and (d), there are only two inflection points from single peaks at the $L_3$ edges (708 eV for Fe and 779 eV for Co) for all of the spectra, and the points are at the same energy. This indicates that both Co and Fe have metallic phases dominantly, i.e., no significant oxidized phase was formed during annealing. It has been reported that vacuum annealing can reduce the Co and Fe oxide phases [28, 29], which is a possible reason of our results. Therefore, the observed decrease in the $M_s$ value is not mainly from the oxidization of CoFeB. It is also noticeable that the intensity and broadness of the peaks in the first-derivative XAS spectra are different from each other, reflecting the fact that the atomic environment of both Co and Fe in each CoFeB layer depends on sort of the underlayer.

The XMCD study exhibits how the electron orbital structures of the CoFeB layers are affected by the atomic environment. Figures 3(a) and (b) present the XMCD spectra for 0° and 70° incident angles, respectively. The trend in the magnitude of the XMCD intensity is almost the same as that in the $M_s$ values (see inset of Figs. 3(a) and (b)). The $m_o$ and the effective spin magnetic moment ($m_s^{eff}$) of Co and Fe atoms in the films can be estimated using the following equations [30,31];

$$-\frac{2}{3}\left(\frac{\Delta A_{L_3} + \Delta A_{L_2}}{A_{total}}\right) \cdot n_h \mu_B = m_o, \text{and} \tag{1}$$

$$-\left(\frac{\Delta A_{L_3} - 2\Delta A_{L_2}}{A_{total}}\right) \cdot n_h \mu_B = m_s^{eff}, \tag{2}$$

where $A_{total} = \int_{L_3} I(E)dE + \int_{L_2} I(E)dE$ is the XAS integral summed over the $L_3$ and $L_2$ edges; $\Delta A_{L_3} = \int_{L_3} \Delta I(E)dE$ and $\Delta A_{L_2} = \int_{L_2} \Delta I(E)dE$ are the integrals of the XMCD spectra at the $L_3$ and $L_2$ edges, respectively [see the inset of Fig. 3 (c)]; $\Delta I = I_+(E) - I_-(E)$, $n_h$ is the hole number of the $d$ band for the transition metals, and $\mu_B$ is the Bohr magneton. Figures 3 (c)–(f) show the XMCD integrals, $\Delta A_{L_3} + \Delta A_{L_2}$. Magnitude of the integrals of both Co and Fe shows the same trend as $M_s$ values in terms of the underlayers.

Figures 4(a) and (b) display the perpendicular $m_o$ ($m_o^\perp$) and the in-plane $m_o$ ($m_o^\parallel$) values, calculated using the relation $m_o(\theta) = m_o^\perp \cos^2\theta + m_o^\parallel \cos^2\theta$, where $\theta$ is the angle between the magnetization and the film normal. In this paper, we use the moment value per single hole (=$m_o/n_h$) rather than the moment value per atom because the $n_h$ values of Co and Fe in CoFeB are difficult to determine precisely in various atomic environments. Using the sum rule, the $m_o/n_h$ value can be directly obtained from the XMCD spectra without calculation of the $n_h$. Therefore, only the trend in $\Delta m_o = m_o^\perp - m_o^\parallel$ per single hole is considered to evaluate the orbital moment anisotropy rather than the atomic magnetic anisotropy energy based on the Bruno model [32]. We find two important observations. First, the trend of change in $\Delta m_o$ of Fe is consistent with that in $K_i$ of the films with respect to the underlayer as shown in Fig. 4 (c), while $\Delta m_o/n_h$ of Co does not follow the trend. In CoFeB/MgO systems, it has been reported that the *p-d* orbital hybridization between Fe and O at the interface is the microscopic origin of the PMA. Therefore, our observation supports that the PMA of the films is mainly governed by the *p-d* hybridization at the CoFeB/MgO interface [33,34]. Second, $m_o^\perp/n_h$ of Fe shows strong underlayer dependence, while $m_o^\parallel/n_h$ is independent with the underlayer. Therefore, atomic intermixing which induces the magnetic deadlayer results in the *d-d* hybridization between Fe and the underlayer elements with the in-plane direction, then suppresses the $m_o^\perp/n_h$ [35]. In case of Co, both $m_o^\perp/n_h$ and $m_o^\parallel/n_h$ strongly depend on the underlayer, i.e. Co orbital structures are isotropically affected by the atomic intermixing. Moreover, $\Delta m_o/n_h$ values of Co are larger than those of Fe in cases of the films with Ta and Ti underlayers. This indicates that Co is also an important ingredient for PMA of CoFeB/MgO systems with Ti and Ta underlayers.

In conclusion, the results of this study demonstrate that the atomic environment in the CoFeB layer is influenced by underlayer materials, resulting in the underlayer dependence of the $M_s$ and PMA. The anisotropy of the $m_o$ also shows strong underlayer dependence. We also manifest that the atomic intermixing between CoFeB and underlayers suppress the $m_o^\perp$ of the Fe, resulting in the weak PMA, while both $m_o^\perp$ and $m_o^\parallel$ of Co are simultaneously suppressed with the intermixing.

Acknowledgements

This work was partly supported by JSPS KAKENHI Grant Numbers 15H05702, 26870300, 26870304, 26103002, 25220604, Collaborative Research Program of the Institute for Chemical Research, Kyoto University, R & D project for ICT Key Technology of MEXT from the Japan Society for the Promotion of Science (JSPS), and the Cooperative Research Project Program of the Research Institute of Electrical Communication, Tohoku University. This work has also been performed with the approval of the SPring-8 Program Advisory Committee (Proposal Nos. 2016A0117, 2016B0117). S. K. was supported from overseas researcher under Postdoctoral Fellowship of Japan Society for the Promotion of Science (Grant Number 2604316). B.-G.P. acknowledges a financial support from National Research Foundation of Korea (NRF-2015M3D1A1070465). K.-J.K acknowledges support from the KAIST start-up funding.

Table I. Film structures with Ta, W, IrMn and Ti underlayers.

| Film structure | $M_s$ (MA/m) | $B_a$ (T) |
|---|---|---|
| Sub./Ta(5)/Ti(5)/CoFeB(1.0)/MgO(1.6)/Ta(2) | 0.47 | 1.04 |
| Sub./Ta(3)/CoFeB(1.0)/MgO(1.6)/Ta(2) | 0.61 | 1.31 |
| Sub./Ta(5)/Ti(5)/IrMn(5)/CoFeB(1.0)/MgO(1.6)/Ta(2) | 0.82 | 1.59 |
| Sub./W(4)/CoFeB(1.0)/MgO(1.6)/Ta(2) | 1.01 | 1.76 |

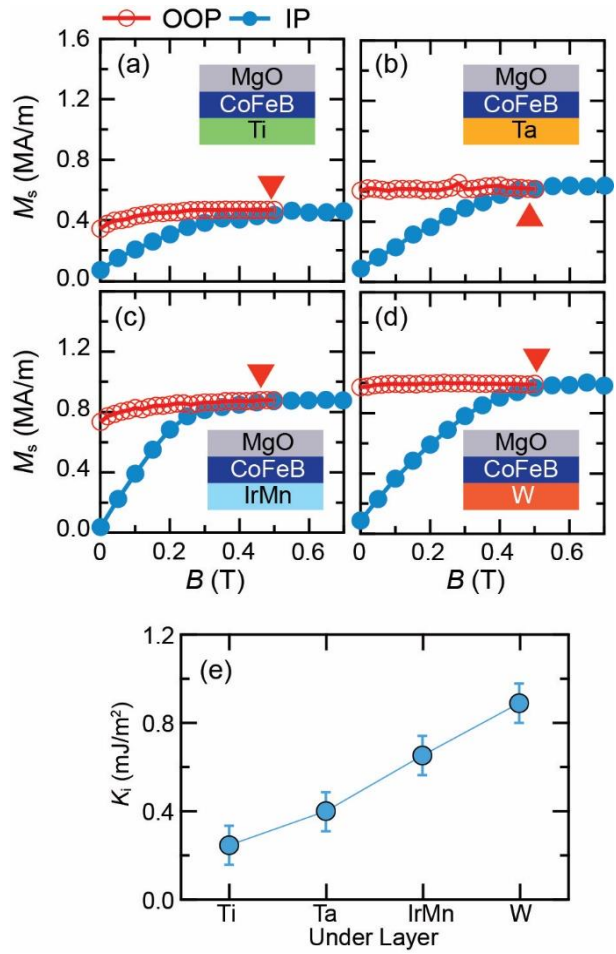

Figure 1. (a)–(d) $M_r$ vs. H curves of the films. The insets show the core structure of each film. Red arrows indicate the magnetization saturation point under the in-plane (hard axis) field. OOP and IP are abbreviations of 'out-of-plane' and 'in-plane', respectively. (e) $K_i$ plots with respect to the underlayer.

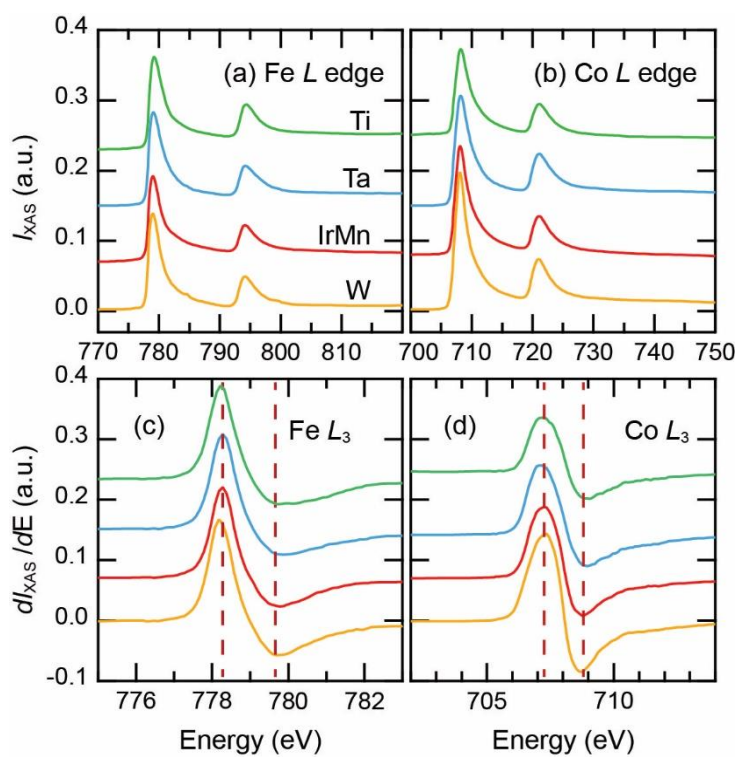

Figure 2. XAS spectra of (a) Fe and (b) Co. (c) and (d) display the first-derivative XAS of the Fe and Co $L_3$ edges, respectively. The dotted red lines indicate the maximum and minimum peaks (inflection points of the XAS spectra) in the spectra.

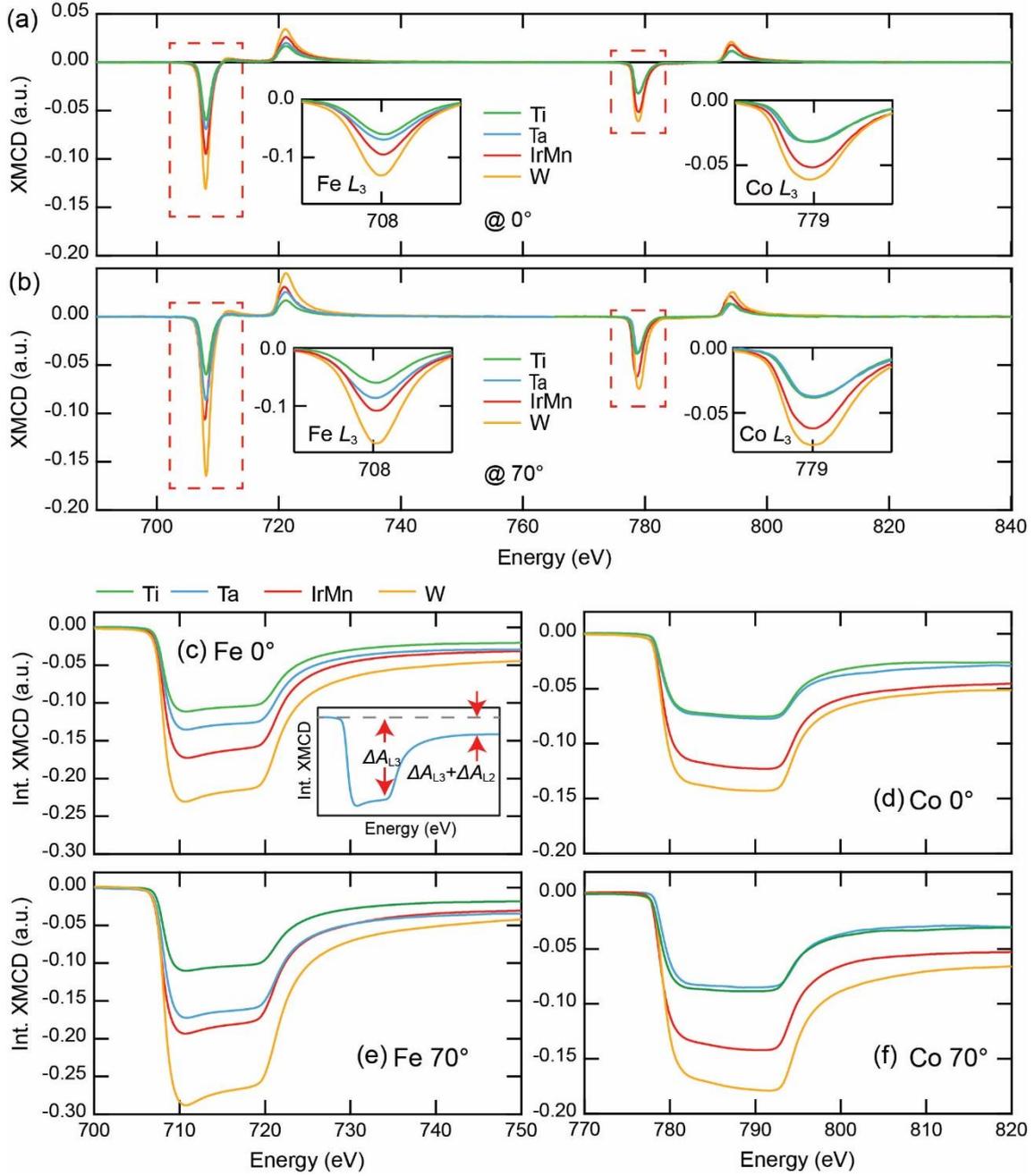

Figure 3. (a) and (b) XMCD spectra and (c) and (d) integral XMCD spectra of the films. Incident X-ray angles with respect to the film normal and elements (Co or Fe) are displayed in the figures. The inset in Fig. (c) illustrates how to determine $\Delta A_{L_3}$ and $\Delta A_{L_3} + \Delta A_{L_2}$ from the integral XMCD spectra.

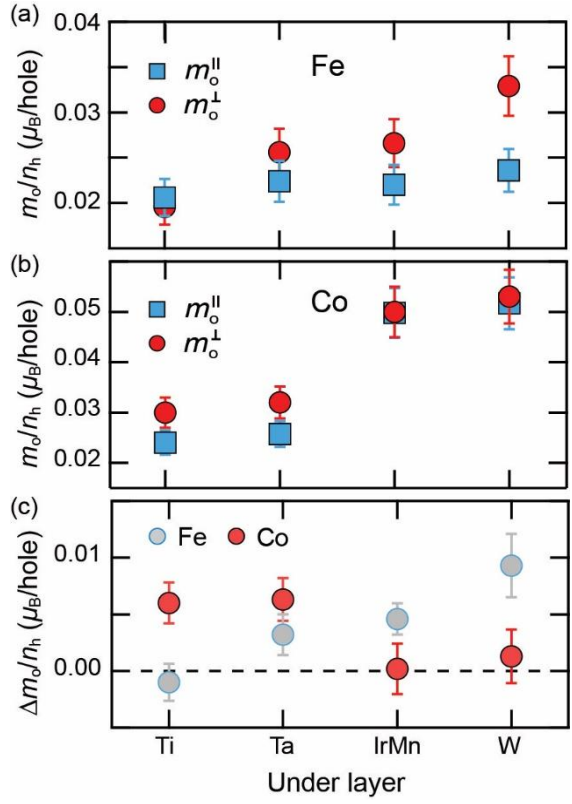

Figure 4. $m_o$ values of (a) Fe and (b) Co, and (c) $\Delta m_o$ values with respect to the underlayers.